# Capacitorless Model of a VO$_2$ Oscillator

**M. A. Belyaev[1], A. A. Velichko[1]**

[1]Institute of Physics and Technology, Petrozavodsk State University, 33, Lenin str., 185910, Petrozavodsk, Russia

E-mail: biomax89@yandex.ru

**Abstract**. We implement a capacitorless model of a VO$_2$ oscillator by introducing into the circuit of a field-effect transistor and a VO$_2$ thermal sensor, which provide negative current feedback with a time delay. We compare the dynamics of current and voltage oscillations on a switch in a circuit with a capacitor and without a capacitor. The oscillation period in the capacitorless model is controlled in a narrow range by changing the distance between the switch and the sensor. The capacitorless model provides the possibility of significant miniaturization of the oscillator circuit, and it is important for the implementation of large arrays of oscillators in oscillatory neural networks to solve the problem of classification and pattern recognition.

## 1. Introduction

A traditional method of computing, based on Boolean logic and implemented using CMOS circuits, suffers from its technological limitations on the productivity growth of computing devices, and, consequently, leads to the limit on data processing speed [1]. Alternative approach to solve this problem offers a radically different way of organizing calculations, based on the dynamics of nonlinear systems [2, 3], that resemble the principles of the human brain operation [4], where billions of neurons experience impulse changes in electrical potential. These systems, called spiking neural networks (SNNs) or third-generation networks, are implemented using various techniques [5–7]. The principles of SNN information processing are based on the analysis of the sequence of pulses: the order of receiving pulses at the network outputs, the distances between pulses and the time of the first appearance of a pulse at any output, as well as on the registration of synchronous activity of different groups of neurons within certain time windows. The latter type of information representation, called neural population coding, reflects the collective dynamics of nonlinear systems, and represents the result of another type of spike networks - impulse oscillatory neural networks (ONNs). Therefore, ONN represents an array of connected oscillators (forced generators or auto-generators), and the principle of adjustment to entire system synchronization [8–10] may underlie the new ways of information processing.

ONNs resemble Hopfield networks [11], where the network dynamics converges to one of the equilibrium positions. However, there are a number of significant differences related to the physics of processes and the presence of complex synchronization effects. The synchronization effect and its metrics are powerful tools for using the collective dynamics of an oscillators array to implement the cognitive functions of a neural network and the information processing. The development of a new element base for oscillators as part of neural networks faces many technological difficulties. And so far, none of the laboratories has been able to produce an oscillatory network with a significant number of elements suitable for processing large amounts of information. The most common electrical circuits of the relaxation oscillator [3,12–14] use a bistable switching element and a capacitance that is charged in a high-resistance state and discharged in a low-resistance state of a bistable element. Switching elements can be implemented based on the memristive switching effect [6,12], magnetic moment transfer [15] and metal-insulator phase transition (MIT) [3,13,14,16]. The use of MIT material allows the thermal interaction between the switches to be used for communication between oscillators [13,16].

One of the technological problems of miniaturization of oscillator circuits is the application of capacitors that occupy a large area of the crystal using standard CMOS technology. For example, with an inter-electrode dielectric thickness of 10 nm, the specific capacitance of the capacitor is ~ 3 fF / $\mu m^2$, and the capacitances with a nominal value of more than 1 pF are needed to generate oscillations on submicron-sized switches [16]. When implementing the capacitor model of the oscillator, nearly 330 $\mu m^2$ of the crystal area is required to manufacture the capacitor, while only ~ 1 $\mu m^2$ is used to manufacture the switch and the load or current resistors. Therefore, the area of the oscillator can be significantly reduced (by hundreds of times), if the capacitance is not used in the oscillator circuit.

The capacitor in the $VO_2$ oscillator circuit has two main functions. First, the capacitor accumulates the energy necessary for heating the switching channel and channel's transition to the metal state. Second, a time delay between pulses is created due to the finite time of charging and discharging the capacitor to threshold voltages. In the absence of a capacitor, the self-oscillations immediately die out, achieving a stable operating point [16].

In the current study, we propose a capacitorless oscillator model, where energy storage and time delays are implemented due to the introduction of a control transistor and an additional $VO_2$-thermal sensor in to the circuit.

## 2. Circuits of VO$_2$ oscillators

The main element of oscillator circuits is a two-electrode switch based on vanadium dioxide. This material has a MIT property with a sharp change in resistivity up to 5 orders of magnitude at a phase transition temperature $T_{th}$ ~ 340 K, which leads to the presence of the electric switching effect in the $VO_2$ switch [17]. We performed the simulation of two-electrode planar structures in the COMSOL Multiphysics environment. The switch was based on $VO_2$ film with a size of 1 μm x 1 μm and a thickness of 100 nm, with the physical parameters presented in [13]. To simplify the model, metal contacts were not used and voltage was applied to the edges of the $VO_2$ film. Sapphire was used as the substrate material, and the substrate was 20 μm x 20 μm x 20 μm in size. The I–V characteristic of the switch was simulated with a ramp voltage $V_{sw}$ (~ $10^3$ V/s) in the range from 0 to 5 V. The I–V characteristic of the switch is presented in Figure 1a. Two competing processes determine the channel temperature: Joule resistive heating by the flowing current and heat dissipation into the environment. As a result, on the I-V characteristic, we can define high-resistance branch ($R_{off}$ ~ 56 kΩ), low-resistance branch ($R_{on}$ ~ 620 Ω), switching on points ($V_{th}$ = 4.7 V, $I_{th}$ = 1.2 · $10^{-4}$ A) and switching off points ($V_h$ = 1.2 V, $I_h$ = 1.5 · $10^{-3}$ A).

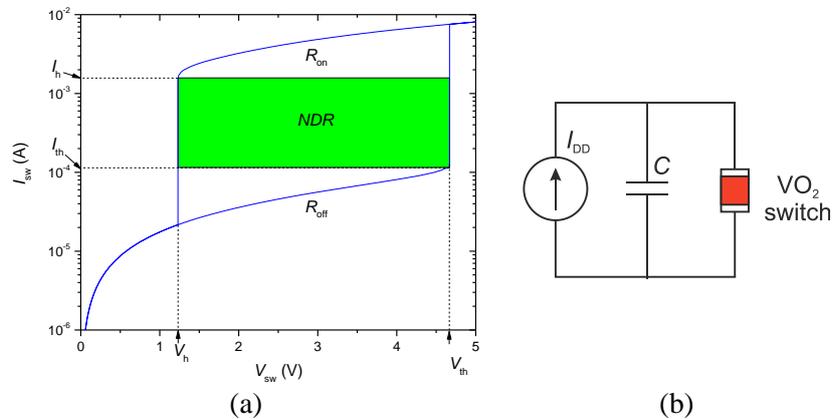

**Figure 1**. Model I-V characteristic of the $VO_2$ switch (a), and a standard oscillator circuit with a capacitor (b).

Figure 1b presents a standard circuit of a VO$_2$ oscillator with a capacitor. Self-oscillations emerge in the circuit, if the $I_{DD}$ supply current falls within the interval between $I_{th}$ and $I_h$ (the portion of the I – V characteristic with negative differential resistance (NDR) is highlighted in green in Figure 1a). The oscillation mode is set in the circuit due to the periodic charging and discharging of the capacitor $C$. When the capacitor is charged to a voltage $V_{th}$, the switch is turned on and its resistance decreases stepwise to $R_{on}$. After that, the capacitance $C$ rapidly discharges through $R_{on}$ until the voltage at the switch decreases to $V_h$, and the switch enters a high-resistance state with resistance $R_{off}$.

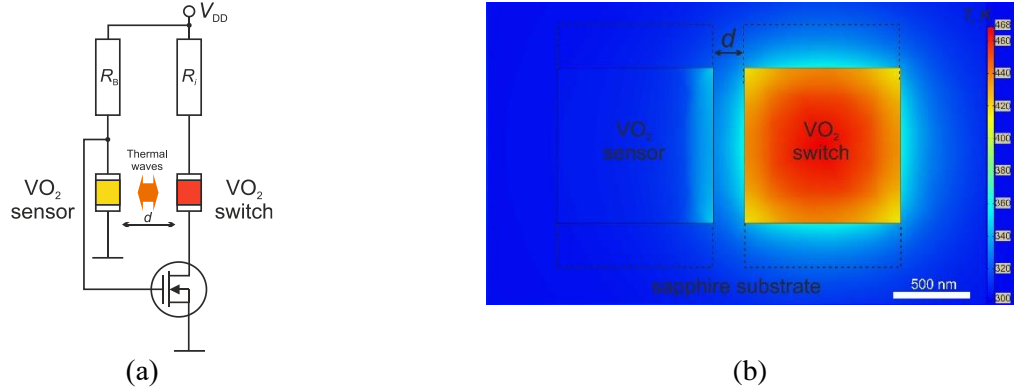

(a)          (b)

**Figure 2.** Circuit of the capacitorless model of the oscillator (a), and an example of the temperature distribution on the main switch and sensor at $d = 200$ nm. The dotted line indicates the region of attachment of the electrodes to the VO$_2$ film, and the white line indicates the scale.

The capacitorless model of the VO$_2$ oscillator is presented in Figure 2a, where a field-effect transistor and a VO$_2$ thermal sensor are added to the circuit. The VO$_2$ sensor repeats the design of the switch and operates only on the high-resistance branch of the I–V characteristic (Figure 1a), with a voltage on the sensor less than $V_{th}$. The switch and the sensor are located in close proximity to each other at a distance $d$, varying within $d = 100$-$700$ nm. The field effect transistor in the circuit has the funciton of a controlled load resistance. The control is performed by applying voltage to the base of the transistor from the VO$_2$ sensor, and the voltage depends on the temperature of the sensor. As the temperature of the sensor changes with a delay relative to the temperature of the switch, it is possible to achieve the effect of complete heating of the switch with its subsequent shutdown. The operating principle of capacitorless model of the oscillator is explained by simplified equivalent circuit presented in Figure 3a, where $R_L$ is the total load resistance of the transistor and limiting resistance $R_i$. The current through the switch is determined by the operating point of the circuit, which position can be found graphically, as the point of intersection of the I – V characteristic of the switch and the load curve $I_{sw} = (V_{DD}-V_{sw}) / R_L$ (Figure 3b). In the initial state, the transistor is open, and the resistance $R_L$ has a small value (for example, $R_L \sim 1.4$ k$\Omega$). When applying voltage $V_{DD} = 5$ V, the intersection of the load curve and the I–V characteristic is located on the low-impedance branch of the I–V characteristic (point "A" in Figure 3b). At this point, a large current $I_{sw}$ flows through the switch, the VO$_2$ channel of the switch heats up, and, subsequently, heats the sensor. When the sensor heats up, its resistance drops. The voltage at the transistor's base decreases, causes the transistor to close and increases $R_L$. An increase in $R_L$ causes a decrease in the inclination angle of the load curve. At some point, the load curve stops crossing the low-resistance branch of the I – V characteristic and begins to cross the high-resistance branch (point "B" in Figure 3b). In this case, the switch goes into off state and the current $I_{sw}$ significantly decreases. The VO$_2$ channel of the switch cools down and the $R_L$ value begins to decrease, increasing the inclination angle of the load curve. The change in $R_L$ has a time delay relative to the change in $I_{sw}$ because of the finite propagation time of thermal waves. It ensures the required heating of the channel and determines the period of self-oscillations of the capacitorless oscillator model.

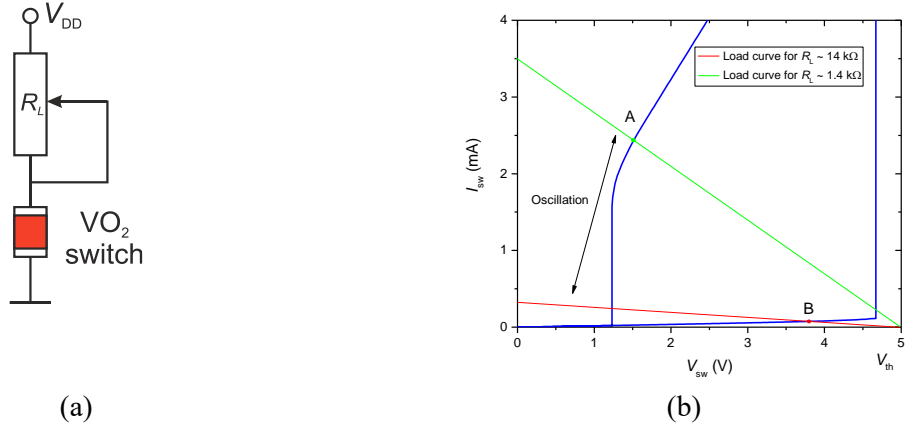

(a)             (b)

**Figure 3**. A simplified circuit of the capacitorless oscillator model (a) and the I–V characteristic of the switch and load curves for different values of the resistance $R_L$ (b).

The main circuit presented in Figure 2a uses a field-effect transistor (n-channel mosfet). The gate of the transistor is supplied with voltage from a divider formed by the $R_B$ resistance and the $VO_2$ resistance of the thermal sensor. Infineon BSC028N06LS3 was selected as the field-effect transistor, and its SPICE model parameters were taken from the library of the LTSpice software. The $R_B$ resistance is selected in such way that the field effect transistor is open at room temperature, and transistor closes, when $VO_2$ sensor is heated. The limiting resistance $R_i$ enables an additional adjustment of the maximum current $I_{sw}$ of the switch. The following circuit parameters were used in the study: $R_B = 60$ kΩ, $R_i = 200$Ω, $V_{DD} = 5$V.

## 3. Modeling of VO₂ oscillator circuits

Figure 4a presents the oscillograms of the current $I_{sw}$ and voltage $V_{sw}$ on the switch in the oscillator model with a capacitor (Figure 1b) at $I_{DD} = 0.6$ mA and $C = 10$ pF. Voltage oscillations with a period of $T \sim 120$ ns occur in the $V_{sw}$ range from $V_h' = 0.5$ V to $V_{th}' = 5$ V. The dynamic threshold characteristics ($V_h'$, $V_{th}'$) differ from the static characteristics ($V_h$, $V_{th}$), and it is usually observed at high frequencies [13]. The charging of the capacitor $C$ is accompanied by an increase in $V_{sw}$ from $V_h'$ to $V_{th}'$, and the discharge of the capacitor is characterized by a sharp drop in $V_{sw}$ from $V_{th}'$ to $V_h'$ and an increase in current $I_{sw}$, observed in Figure 4a.

Figure 4b presents the oscillograms of the current $I_{sw}$ and voltage $V_{sw}$ on the switch and the time dependences of the resistance change of the thermal sensor $R_s$ in the capacitorless oscillator model. The distance between the switch and the sensor was $d = 200$ nm. The oscillogram $V_{sw}$ has a more complex form than in the model with a capacitor, and for its interpretation on the graph, there are digital designations of areas. Section "1" corresponds to the off state of the switch ($I_{sw} \sim 0$ mA), when the voltage $V_{sw}$ increases from 0 to $V_{th}'$. The transistor in the section "1" goes from closed to open state, as the resistance of the sensor increases because of its cooling. After reaching the threshold voltage $V_{th}'$, the switch turns on and the circuit goes to a stable operating point on the low-resistance branch of the I – V characteristic of the switch (section "2" in Figure 4b). In this case, the current through the switch $I_{sw}$ increases significantly. At this moment, the channel and the substrate in the switch area begin to heat up intensively due to Joule heating (section "3"). As the sensor is located on the substrate at a distance of $d = 200$ nm, it begins to heat up with a delay determined by the propagation time of thermal waves. When the sensor heats up, its resistance begins to fall (Figure 4b). This causes a decrease in the voltage at the gate of the field-effect transistor and leads to transistor closure (section "4"). As a result, the switch also goes into the off state. The temperature distribution at the switch and at the sensor is presented in

Figure 2b. After the sensor resistance decreases significantly, the field transistor is completely closed (section "5"), and the voltage $V_{sw}$ equals to 0 V. The duration of this section corresponds to the time the sensor cools down.

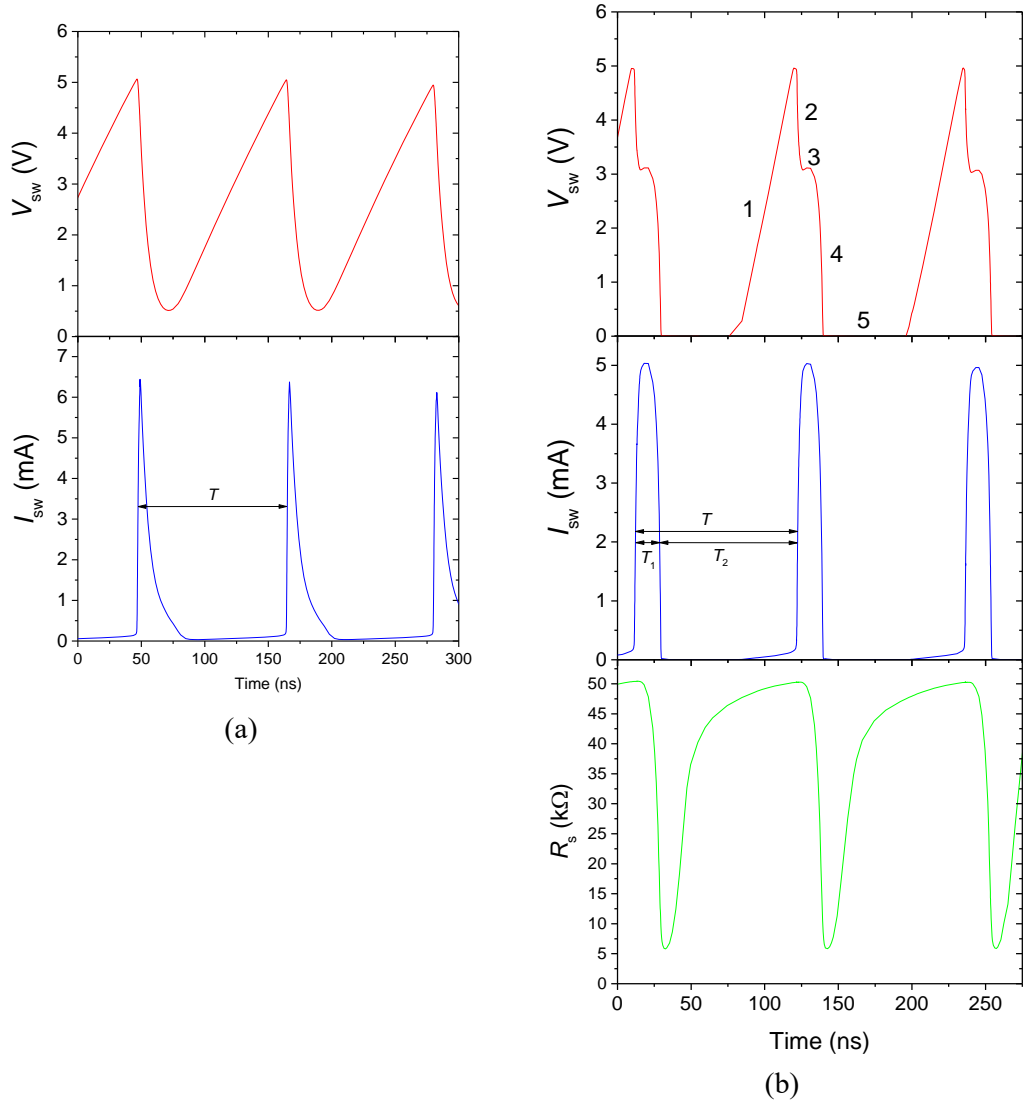

**Figure 4.** Oscillograms of the voltage $V_{sw}$ and current $I_{sw}$ of the switch in the circuit with a capacitor (a) and in a capacitorless model (b).

The oscillogram of the current $I_{sw}$ reflects the oscillation period $T \sim 110$ ns; it consists of the delay time $T_1 \sim 15$ ns required for heating the sensor after the switch is turned on and the sensor cooling time $T_2 \sim 95$ ns. In this model, the oscillation period can be controlled by changing the distance $d$ (Figure 5a). As $d$ increases from 100 to 700 nm, the oscillation period $T$ increases from $\sim 100$ to $\sim 170$ ns. The increase in $T(d)$ is mainly caused by an increase in the delay of $T_1$ from 10 to 60 ns. The thermal waves require more time to travel a longer distance $d$, or, in other words, the heating of a larger area takes longer. The cooling time $T_2$ increases insignificantly, as the cooling time is mainly determined by the maximum heating temperature of the switch channel (amplitude of the heat wave), which depends on the heating power of the channel.

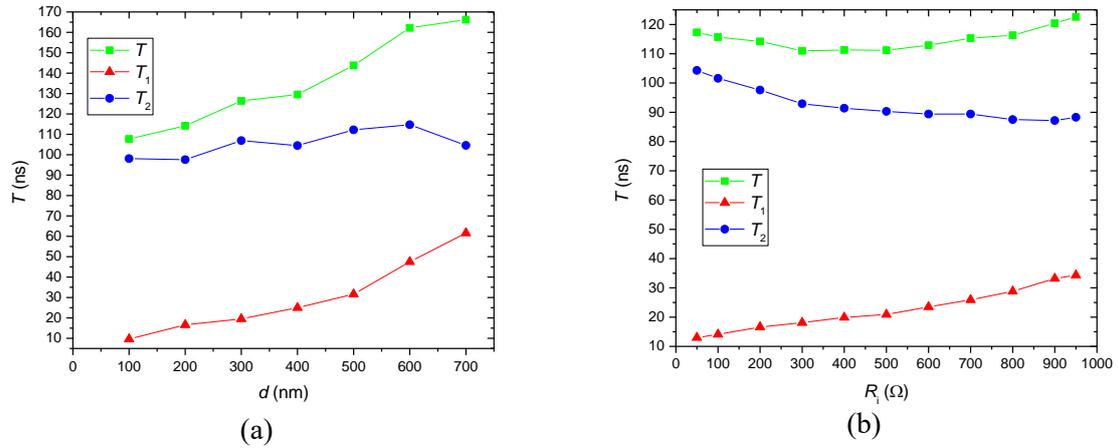

**Figure 5.** The dependence of the oscillation period $T$ and the times $T_1$ and $T_2$ on a change in the distance $d$ (a) and on a variation in the current-limiting resistance $R_i$ (b).

The simulation results of the period $T$ dependence on the value of current-limiting resistor $R_i$ in the range from 50 to 950 Ω are presented in Figure 5b. The oscillation period is practically independent of the resistance $R_i$, because despite the increase in the sensor heating time $T_1$ from 13 to 34 ns, its cooling time $T_2$ decreases from 104 to 88 ns. The decrease in $T_2$ is explained by the fact that at large $R_i$ the channel heats up to a lower temperature.

### 4. Conclusion
In the current study, we present the dynamics of oscillations in the circuits with a capacitor and without a capacitor. The $I_{sw}$ and $V_{sw}$ oscillograms of the two circuits are significantly different due to the difference in the mechanisms of energy storage and the time delay between switching. The control of the oscillation period is much simpler in the circuit with a capacitor. Whereas in the capacitorless model, the control of the oscillation period occurs in a narrow range by changing the distance between the switch and the sensor. The change in the period of oscillations can also be accomplished by using the circuit parameters (voltage $V_{DD}$, parameters of a field-effect transistor), which influence remains to be analyzed in the future studies. The most important advantage of the capacitorless model is the possibility of the oscillator miniaturization: the capacitor model requires about 330 μm$^2$ of silicon area, while the presented capacitorless model would require less than 5 μm$^2$, taking into account the transistor area. This characteristic is especially important for the implementation of large oscillator arrays that can be used to manufacture oscillatory neural networks [18] for solving classification and pattern recognition problems.

**Acknowledgment**
The authors express their gratitude to Dr. Andrei Rikkiev for the valuable comments in the course of the article translation and revision.

**References**
[1] Bernstein K, Cavin R K, Porod W, Seabaugh A and Welser J 2010 Device and architecture outlook for beyond CMOS switches *Proceedings of the IEEE* vol 98 pp 2169–84
[2] Bhanja S, Karunaratne D K, Panchumarthy R, Rajaram S and Sarkar S 2016 Non-Boolean computing with nanomagnets for computer vision applications *Nat. Nanotechnol.* **11** 177–83
[3] Kumar S, Strachan J P and Williams R S 2017 Chaotic dynamics in nanoscale NbO2 Mott memristors for analogue computing *Nature* **548** 318–21
[4] McKenna T M, McMullen T A and Shlesinger M F 1994 The brain as a dynamic physical system. *Neuroscience* **60** 587–605

[5]     Pani D, Meloni P, Tuveri G, Palumbo F, Massobrio P and Raffo L 2017 An FPGA Platform for Real-Time Simulation of Spiking Neuronal Networks *Front. Neurosci.* **11** 90

[6]     Adamatzky A and Chua L 2014 *Memristor Networks* (Cham: Springer International Publishing)

[7]     Furber S 2016 Large-scale neuromorphic computing systems *J. Neural Eng.* **13** 051001

[8]     Zanin M, Del Pozo F and Boccaletti S 2011 Computation Emerges from Adaptive Synchronization of Networking Neurons ed E Ben-Jacob *PLoS One* **6** e26467

[9]     Malagarriga D, García-Vellisca M A, Villa A E P, Buldú J M, García-Ojalvo J and Pons A J 2015 Synchronization-based computation through networks of coupled oscillators *Front. Neurosci.* **9** 97

[10]    Cosp J, Madrenas J, Alarcón E, Vidal E and Villar G 2004 Synchronization of nonlinear electronic oscillators for neural computation *IEEE Trans. Neural Networks* **15** 1315–27

[11]    Wittek P 2014 *Quantum Machine Learning: What Quantum Computing Means to Data Mining* (Elsevier Inc.)

[12]    Wang Z, Joshi S, Savel'ev S, Song W, Midya R, Li Y, Rao M, Yan P, Asapu S, Zhuo Y, Jiang H, Lin P, Li C, Yoon J H, Upadhyay N K, Zhang J, Hu M, Strachan J P, Barnell M, Wu Q, Wu H, Williams R S, Xia Q and Yang J J 2018 Fully memristive neural networks for pattern classification with unsupervised learning *Nat. Electron.* **1** 137–45

[13]    Velichko A, Belyaev M, Putrolaynen V, Perminov V and Pergament A 2018 Thermal coupling and effect of subharmonic synchronization in a system of two VO 2 based oscillators *Solid. State. Electron.* **141** 40–9

[14]    Shukla N, Parihar A, Freeman E, Paik H, Stone G, Narayanan V, Wen H, Cai Z, Gopalan V, Engel-Herbert R, Schlom D G, Raychowdhury A and Datta S 2015 Synchronized charge oscillations in correlated electron systems *Sci. Rep.* **4** 4964

[15]    Locatelli N, Cros V and Grollier J 2014 Spin-torque building blocks *Nat. Mater.* **13** 11–20

[16]    Pergament A, Velichko A, Belyaev M and Putrolaynen V 2018 Electrical switching and oscillations in vanadium dioxide *Phys. B Condens. Matter* **536** 239–48

[17]    Belyaev M A, Boriskov P P, Velichko A A, Pergament A L, Putrolainen V V., Ryabokon' D V., Stefanovich G B, Sysun V I and Khanin S D 2018 Switching Channel Development Dynamics in Planar Structures on the Basis of Vanadium Dioxide *Phys. Solid State* **60** 447–56

[18]    Velichko A, Belyaev M, Putrolaynen V and Boriskov P 2018 A New Method of the Pattern Storage and Recognition in Oscillatory Neural Networks Based on Resistive Switches *Electronics* **7** 266